\documentclass[a4paper,twocolumn,11pt,unpublished]{quantumarticle}
\pdfoutput=1
\usepackage[utf8]{inputenc}
\usepackage[english]{babel}
\usepackage[T1]{fontenc}
\usepackage{amsmath,bm,amsfonts}
\usepackage{amsmath,bm,amsfonts}
\usepackage{mathtools}
\usepackage{physics}

\usepackage{graphicx}
\usepackage{hyperref}

\usepackage[numbers,sort&compress]{natbib}
\usepackage[dvipsnames]{xcolor}
\newcommand{\fix}[1]{#1}

\newtheorem{theorem}{Theorem}[section]

\newtheorem{lemma}[theorem]{Lemma}

\begin{document}

\title{Absorbing state transitions with discrete symmetries}

\author{Hyunsoo Ha}
\affiliation{Department of Physics, Princeton University, Princeton, NJ 08544, USA}
\orcid{0000-0002-3288-9241}
\author{David A. Huse}
\affiliation{Department of Physics, Princeton University, Princeton, NJ 08544, USA}
\orcid{0000-0003-1008-5178}
\author{Rhine Samajdar}
\affiliation{Department of Physics, Princeton University, Princeton, NJ 08544, USA}
\affiliation{Princeton Center for Theoretical Science, Princeton University, Princeton, NJ 08544, USA}
\orcid{0000-0001-5171-7798}
\maketitle

\begin{abstract}
 Robust phases of matter, which remain stable under small perturbations, are of fundamental importance in statistical physics and quantum information. Recent advances in interactive quantum dynamics have led to renewed interest in out-of-equilibrium dynamical phases and associated phase transitions in both classical and quantum many-body systems. Motivated by these developments, we investigate whether a stable absorbing phase can exist in one-dimensional classical stochastic systems, with local update rules, in the presence of fluctuations. We study models with multiple absorbing states related by discrete symmetries, such as $\mathbb{Z}_2$ for two-state systems, and $\mathbb{Z}_3$ or $\mathbb{S}_3$ for three-state systems. In these models, domain walls perform random walks and coarsen under local rules, which, if perfect, eventually bring the system to an absorbing state in polynomial time. However, imperfect feedback can cause domain walls to branch, potentially leading to an opposing active phase. While two-state models exhibit a well-known transition between absorbing and active phases as the branching rate increases, in three-state models with only local dynamics, branching is a relevant perturbation, ruling out a robust absorbing phase under purely local rules. However, we discover that by incorporating nonlocal information into the feedback, the absorbing phase can be stabilized, with the transition between the active and absorbing phases belonging to a new universality class. Finally, we outline how these classical rules can be implemented using deterministic quantum circuits and discuss their connections to passive error correction.
\end{abstract}

The classical and quantum dynamics of stochastic processes provide a rich playground to study universal phenomena in out-of-equilibrium settings~\cite{odor2004_nonequilibrium_review}.\,A classic example of such an \textit{intrinsically} nonequilibrium phenomenon is the phase transition of a dynamical many-body system to an absorbing state.  An absorbing state is a stable state of the system's dynamics that the dynamics may make the system go to, but which the system cannot leave. Depending on the strength of fluctuations, the system may or may not reach an absorbing state in polynomial time; in the former (latter) regime, the system is said to be in an absorbing (active) \textit{phase}. These two dynamical phases are separated by an absorbing state phase transition~\cite{Hinrichsen_absorbing_review,henkel_absorbing_book}. While extensively studied for classical systems, recently, such transitions have acquired newfound significance for interactive quantum dynamics~\cite{Carollo_Lesanovsky_2019,Gillman_Lesanovsky_2021,Carollo_Lesanovsky_2022,Gillman_Lesanovsky_2022pre,Carollo_Lesanovsky_2022prb}, in which feedback based on measurement outcomes is applied to the quantum system~\cite{marcuzzi_lesanovsky_absorbing+quantum,absorbing+quantum_exp,Lesanovsky_garrahan_QCA+absorbing,Chertkov_2023_absorbingexperiment,Iadecola_Wilson_absorbing+MIPT,buchold_diehl_absorbing+MIPT,sierant_turkeshi_absorbing+mipt,piroli_nahum_absorbing+mipt,friedman_lucas_feedbackerrorcorrection,friedman_nandkishore_absorbing+mipt,han_chen_PCQCA,ravindranath_chen_PCuniversality} to prepare or recover a desired quantum state in the presence of external noise~\cite{kraus_zoller_steerMarkov,roy_gefen_steering,zhou_lukin_steer,odea_khemani_absorbingstate,ravindranath_chen_absorbing+freefermion,chirame_prem_SPT+absorbing,chirame_Burnell_absorbing2topo}.

In a generic nonequilibrium process described by local dynamical rules and featuring a single absorbing state, such a transition typically belongs to the  universality class of directed percolation (DP)~\cite{Janssen1981_DPconjecture,Grassberger1982_DPconjecture}.  However, when the possible absorbing states are not unique~\cite{Grassberger_1984,Grassberger_1989}---say, due to the presence of additional symmetries---the system may exhibit fundamentally distinct classes of transitions. For instance, in a case with two absorbing states related by a global $\mathbb{Z}_2$ symmetry, the transition is known to belong to  the $\mathbb{Z}_2$-symmetric directed percolation (DP2) universality class, also known as the directed Ising class~\cite{Hinrichsen_absorbing_review,odor2004_nonequilibrium_review}.

Absorbing phases with multiple absorbing states are of particular relevance to the dynamics of quantum circuits. Specifically, as we show, quantum circuits with dynamics equivalent to classical reaction-diffusion models with discrete symmetries can drive a quantum state to a \textit{superposition} of absorbing states. Thus, rather than absorbing to single states, the dynamics can host an absorbing \textit{manifold} (e.g., a Bloch sphere for the $\mathbb{Z}_2$-symmetric model). 

In this work, we investigate the existence of absorbing phases and the associated phase transitions in (1+1)-dimensional systems with discrete symmetries.
The simplest example is a chain of qubits with the manifold of (pure and mixed) absorbing states spanned by the two states $|\uparrow\uparrow...\uparrow\rangle$ and $|\downarrow\downarrow...\downarrow\rangle$ that are related by a global $\mathbb{Z}_2$ symmetry. A ``domain wall'', which is simply an up spin adjacent to a down spin, is a point-like object on this chain, which we identify as a particle, $W$. Importantly, under any local updates in the bulk of the chain, the parity of the total number of such domain walls is conserved, reflecting the $\mathbb{Z}_2$ symmetry. 
The dynamics of the system under (deterministic) local quantum channels, which can be interpreted as (stochastic) local measurement and feedback, can therefore be mapped to parity-conserving branching and annihilating random walkers~\cite{zhong_avraham_BAWE}, corresponding to the ``reactions''
$W \xrightarrow[\lambda]{} 3W$ and $2W \xrightarrow[\alpha]{} \emptyset$, where $\lambda$ and $\alpha$ specify the rates of branching and annihilation, respectively.  The absorbing states are those with no domain walls.  Depending on $\lambda$ and $\alpha$, the system may either reach an absorbing state within a time that is polynomial in the system length or remain in active states for an exponentially long time. Here, we present a quantum model with two-site gates that exhibits an absorbing state transition between these two phases that belongs to the parity-conserving~\cite{Tauber_Cardy_prlPC,Tauber_Cardy_PC} universality class (equivalent to DP2 for (1+1)D). 

We also generalize the local measurement and feedback rules to $Q$-state models with $Q$\,$>$\,$2$, examples of which include Potts or clock models, as well as qudits in quantum systems, where the manifold of absorbing states is spanned by $Q$ orthogonal many-body states.
Previous studies have observed that for $Q=3$, a nonzero domain wall branching rate generically leads to an active phase with no absorbing transition \cite{Hinrichsen_1997,Hooyberghs_generalNabsoring1,Lipowski_generalNabsorbing2}. We confirm this behavior and importantly, provide a physical argument to explain it: in models with $Q>2$, typical branching~\cite{fine_tuned} acts as a relevant perturbation to the absorbing phase and inevitably drives the system into the active phase. Additionally, we demonstrate that while {\it local} feedback cannot stabilize the absorbing phase against branching for these $Q>2$ models, it is possible to steer the system towards the absorbing phase.  We do so by utilizing nonlocal information and introducing biases in the updates, in analogy to using {\it nonlocal} information about syndromes for quantum error correction with local operations and (nonlocal) classical communication (LOCC), and we demonstrate this for 3-state models.

\begin{figure}
    \centering
    \includegraphics[width=\linewidth]{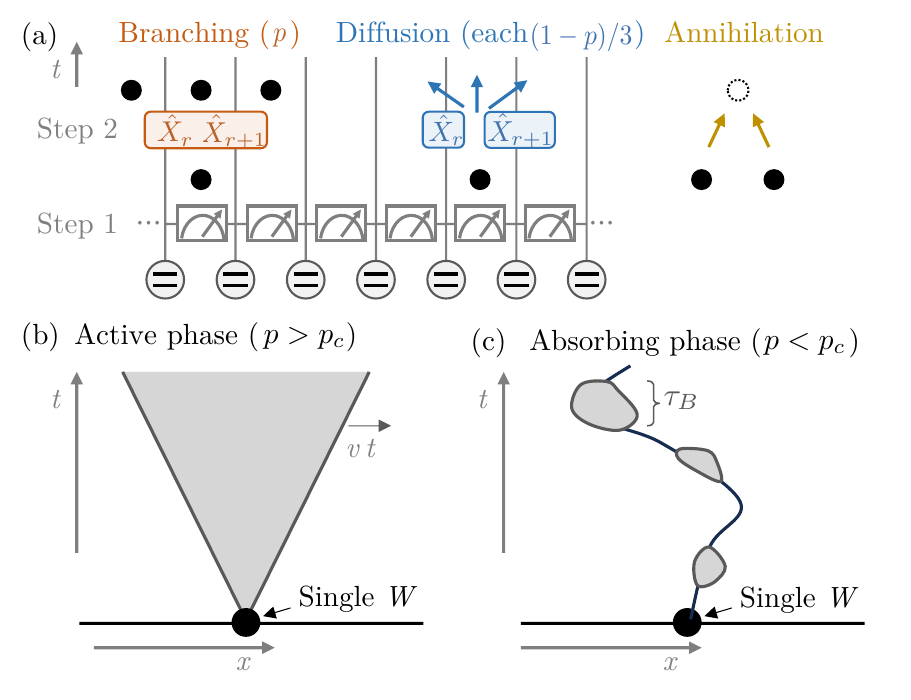}
    \caption{(a) Depiction of a local $\mathbb{Z}_2$-symmetric model with two-site gates that exhibits an absorbing state transition. The channel first measures the presence or absence of a domain wall ($\hat{Z}_r\hat{Z}_{r+1}$) on each bond and then applies feedback that branches (by acting with $\hat{X}_r\hat{X}_{r+1}$) or diffuses (via $\hat{X}_r$, $\hat{X}_{r+1}$, or $\hat{\mathbf{1}}_r\hat{\mathbf{1}}_{r+1}$) the domain wall if it is present. (b) Schematic representation of the active phase, in which the active domain grows ballistically over time after initializing with only one domain wall.
    (c) Illustration of the absorbing phase, where the dynamics of a single domain wall can be coarse-grained into multi-domain-wall ``bubbles''.
 }
    \label{fig:Z2}
\end{figure}

\section{$\mathbb{Z}_2$-symmetric local models.}
Previous numerical works have established the absence of an absorbing state transition for $\alpha/\lambda$\,$=$\,$\infty$, as the system always stays in the absorbing phase~\cite{Takayasu_Tretyakov_notransition,DP2_notransition1,DP2_notransition2,DP2_notransition3}. A transition can occur for finite $\alpha/\lambda$; previously examined models with this transition have gates acting on three or more sites~\cite{zhong_avraham_BAWE,ravindranath_chen_PCuniversality,han_chen_PCQCA, Grassberger_2013}. Here, we introduce and study a model with \textit{two}-site gates, which does host the absorbing state transition. We use this simple model to revisit the DP2 critical point and investigate properties of both the absorbing and active phases.

We consider a discrete-time classical update rule on a one-dimensional lattice, where particles are the domain walls between neighboring sites. At every time step, each particle is updated \textit{simultaneously}. With probability $p$, a particle branches by creating two new particles on the adjacent bonds. With the remaining probability $1 - p$, it performs a symmetric random walk, moving one bond to the left, staying in place, or moving one bond to the right, each with equal probability. After all updates have been applied, we retain only the parity of the number of particles on each bond, so that pairs of particles on the same bond annihilate.

These classical dynamics can be realized using a \textit{deterministic} quantum channel, as illustrated in Fig.~\ref{fig:Z2}(a). Each discrete time step consists of two layers. In the first layer, a two-site quantum channel effectively records whether a domain wall is present into a classical register on each bond—analogous to performing a projective measurement of $\hat{Z}_r \hat{Z}_{r+1}$. In the second layer, a feedback operation acts conditionally, based on this classical information: if no domain wall is recorded, the state is acted upon by the identity; if a domain wall is recorded, the channel simultaneously implements branching (flipping both neighboring spins) with weight $p$ and a random walk (flipping either one spin or none) with each outcome having equal weight summing to $1-p$. As in the classical case, pairs of domain walls occupying the same bond annihilate, by construction.

The quantum circuit described above implements the classical particle dynamics introduced earlier, and in particular, we numerically investigate the absorbing state transition in an infinitely long system starting with a single initial wall~\footnote{an odd number of domain walls could be prepared by considering infinite boundary conditions with the left and right ends being different domains, or by considering  antiperiodic boundary conditions for a finite system size}. As argued earlier, the domain wall parity is conserved, so at least one $W$ is always present for this choice of initial state. We monitor the number of domain walls $N_{dw}(t)$, and from the scaling assumption \cite{odor2004_nonequilibrium_review,Hinrichsen_absorbing_review}
\begin{align}
    N_{dw}(t) \sim t^{\theta} f\left[ (p-p_c)t^{1/\nu_{\parallel}} \right],
\end{align}
where $f$ is some universal function, we find $p_c=0.24\pm0.01$ and critical exponents consistent with the DP2 class: $\theta=0.29\pm0.01$, and $\nu_\parallel = 3.4\pm0.2$ \cite{Hinrichsen_absorbing_review,odor2004_nonequilibrium_review,Park_2020_DP2beta,supp_mat}.

In the active phase, i.e., $p>p_c$, branching dominates mutual annihilation. Hence, any defect---even a single domain wall---produces an active domain, inside which the (nonzero) number density of walls is spatially uniform on average. The width of this active domain grows ballistically over time as depicted in Fig.~\ref{fig:Z2}(b) (see  SM~\cite{supp_mat} for numerical data). 

On the other hand, for small branching rates $p < p_c$, the system is in the absorbing phase. The dynamics in this phase are governed by the annihilation fixed point ($\lambda=0$)~\cite{Lee_absorbing_fixed_point}, so the branching is irrelevant in the renormalization-group sense. Accordingly, if one starts with a random initial state, the number of walls decays over time as $\sim t^{-1/2}$ via coarsening dynamics driven by pair annihilation~\cite{Tauber_Cardy_prlPC, Tauber_Cardy_PC}. For the single-domain-wall initial state and boundary conditions that we consider, the particle number is odd, so it can never decay to zero. This single initial $W$ particle does not produce an active domain, unlike in the active phase. When the particle does branch, the resultant multiparticle state eventually returns to a single-particle state.  We can thus define a ``bubble'' lifetime $\tau_B$ as the duration of this multiparticle state. The long-time cumulative distribution of the bubble lifetime asymptotically scales as $P_s(\tau_B > t) \sim t^{-3/2}$~\cite{supp_mat}, which corresponds to the survival probability of three particles that perform random walks and are subject to pairwise annihilation on contact. The resulting mean lifetime is of order one, but with this weak power-law tail to long lifetimes.  Hence, these bubbles become dilute when the domain-wall dynamics in the absorbing state are coarse-grained [Fig.~\ref{fig:Z2}(c)].

\section{Three-state models with local feedback.}
We now ask whether a robust absorbing phase can also exist in three-state systems, where each site carries one of three values: 0, 1, or 2. One key difference from the $Q$\,$=$\,$2$ case above is that for $Q$\,$=$\,$3$, there exist two distinct types of domain walls; we label these ``particles'' $R$ ($L$) if the value of the spin increases (decreases) by one unit, modulo 3, on moving rightwards by one site.  Once again, we consider dynamics with discrete symmetries by requiring now that the difference between the total number of $R$ ($n_R$) and $L$ domain walls ($n_L$) is conserved modulo 3:
\begin{align}
n_R(t) - n_L(t) = \text{constant} \quad (\text{mod}~3)~.
\label{eq:nR-nL(3)}
\end{align}
When the microscopic rules are invariant under swapping $R$ and $L$, the model possesses an $\mathbb{S}_3$ symmetry, generated by the permutation of the values 0, 1, and 2. If the evolution rules for $R$ and $L$ differ, however, the associated symmetry is $\mathbb{Z}_3$, corresponding to only cyclic permutations. In either case, as a consequence of this discrete symmetry, the set of all (pure or mixed) absorbing states is spanned by the three many-body states $|00...0\rangle$,  $|11...1\rangle$, and $|22...2\rangle$. 

In this work, we consider a diffusion-reaction model for these two types of domain walls, which could be produced by two-site quantum channels, similar to the two-state model presented above.  Suppose each particle performs a random walk with diffusivity $D_R$ or $D_L$, depending on its species, $R$ or $L$. The minimal reaction rules respecting the constraint given by Eq.~\eqref{eq:nR-nL(3)} are as follows.
The total particle number can decrease by pair annihilation when two different types of walls meet, or by coagulation when two walls of the same type meet and form a single wall of the opposite type:
\begin{align}
    R + L \xrightarrow[\alpha]{} \emptyset~, \, R + R \xrightarrow[\gamma_{R}]{} L~,   \,  L + L \xrightarrow[\gamma_{L}]{} R~;
    \label{eq:3state_decrease}
\end{align}
here, $\alpha$ is the annihilation rate, and $\gamma_{R}$ and $\gamma_L$ are the coagulation rates. Likewise, the simplest branching processes are given by:
\begin{align}
    R\xrightarrow[\lambda_R]{} L + L~, \quad\quad L\xrightarrow[\lambda_L]{} R + R~,
    \label{eq:3state_branching}
\end{align}
where $\lambda_R$ and $\lambda_L$ are the branching rates. Of course, in the $\mathbb{S}_3$-symmetric case, $D_R$\,$=$\,$D_L$, $\gamma_R$\,$=$\,$\gamma_L$, and $\lambda_R$\,$=$\,$\lambda_L$.
Note that any allowed higher-order branching processes (e.g., $R\rightarrow R R L$) will be produced from these elementary processes acting over time. 

Interestingly, in our numerics, we find that this model is always in the active phase if the probability of the branching process is \fix{nonzero~\cite{fine_tuned}}. 
This phenomenology also holds for three-state models in which the annihilation and coagulation are enhanced by using three-site gates~\cite{supp_mat}. Underpinning this observation is the fact that even an infinitesimally small branching process necessarily drives the system into the active phase, as we now outline.

\begin{figure}
    \centering
    \includegraphics[width=\linewidth]{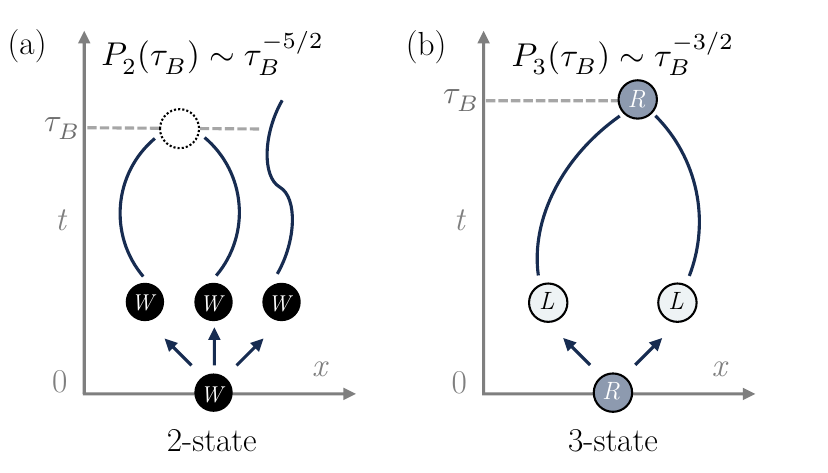}
    \caption{Minimal ``bubble'' 
    for (a) the $\mathbb{Z}_2$-symmetric two-state model, and (b) the $\mathbb{S}_3$- or $\mathbb{Z}_3$-symmetric three-state model. After a single branching event, the probability distribution of the bubble lifetime $\tau_B$ scales as $\tau_B^{-5/2}$ and $\tau_B^{-3/2}$ for the two-state and  three-state local models, respectively. }
    \label{fig:S3_cartoon}
\end{figure}

\begin{figure*}[tb]
    \centering
    \includegraphics[width=\linewidth]{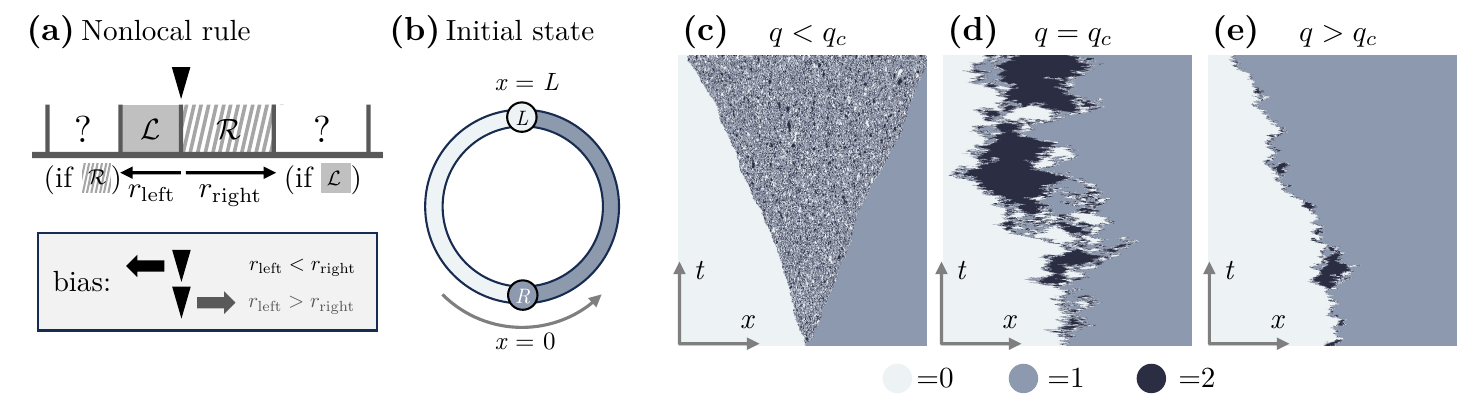}
    \caption{
    (a) Schematic of update rules using nonlocal information and local gates.  The domain wall of interest separates a ``filled'' domain $\mathcal{L}$ on the left from a ``striped'' domain  $\mathcal{R}$ on the right. 
    We examine the next-nearest domain on each side: if the next left domain is striped, we record the distance $r_{\mathrm{left}}$, and if the next right domain is filled, we record $r_{\mathrm{right}}$. If the next-nearest domain does not match the desired type, the corresponding distance is recorded as infinity. The biased walk is then directed towards the side with the shorter distance.
    (b) To avoid boundary effects, we use periodic boundary conditions with two walls separated by a distance larger than any characteristic length scale achievable within our runtime.
    Representative spatial configurations of domains as a function of time are shown (c) in the active phase ($q<q_c$), (d) at the critical point ($q=q_c$), and (e) in the absorbing phase ($q>q_c$).
 }
    \label{fig:nonlocal}
\end{figure*}

\section{Relevance of branching.}
As for the two-state models above, the dynamics initiated from a single domain wall in the absorbing phase can be viewed in terms of bubbles: clusters of domain walls that form when a single particle branches into multiple ones which eventually collapse back to a single-particle state at time $\tau_B$. To test whether branching is a relevant perturbation at the absorbing fixed point, we consider the dynamics initiated by a \textit{single branching event} from a single domain wall, without allowing any additional branching, and analyze the time it takes for the system to return to a single particle. The stability of the absorbing phase depends on whether or not the mean of this bubble lifetime is finite.

In the two-state models, a single particle splits into three upon branching. The bubble ends when two adjacent particles recombine and annihilate.
As shown in the SM~\cite{supp_mat}, the probability distribution of $\tau_B$ scales as $P_{2}(\tau_B)\sim\tau_B ^{-5/2}$ for late times. Consequently, the average bubble lifetime $\langle \tau_B \rangle_{2}$ is finite:
\begin{align*}
    \langle \tau_B \rangle_{2} = \int_{t_0}^\infty \tau_B P_2(\tau_B) d\tau_B \sim \int_{t_0}^\infty \tau_B^{-3/2} d\tau_B < \infty,
\end{align*}
where $t_0$ is determined by microscopic details. Therefore, the branching events are irrelevant, and the absorbing phase may survive for small but nonzero branching probabilities (as argued by~\citet{Tauber_Cardy_PC}).

However, in the three-state model, the branching rules are fundamentally different: the minimal branching event is a single domain wall ($R$) splitting into two walls of the opposite type ($2L$) [see Fig.~\ref{fig:S3_cartoon}(b)].
In this case, the probability distribution of the bubble lifetime for two randomly walking particles to recombine at time $\tau_B$ asymptotically goes as $P_3(\tau_B)\sim\tau_B^{-3/2}$, leading to a divergent average lifetime $\langle \tau_B \rangle_{3}$:
\begin{align*}
    \langle \tau_B \rangle_{3} = \int_{t_0}^\infty \tau_B P_3(\tau_B) d\tau_B = \int_{t_0}^\infty \tau_B^{-1/2} d\tau_B \rightarrow \infty.
\end{align*}
As a result, the fraction of time the particle spends as a bubble goes to one, even if the branching rate is small.  This is even without considering that each particle within the bubble can branch further.  Thus, the branching becomes a relevant perturbation to the absorbing fixed point for $Q\geq 3$ and drives the system into the active phase.

\section{Transition induced by nonlocal information.}
Even though the simple three-state model above does not generically host an absorbing phase, one can facilitate an absorbing state transition via nonlocal actions. We note that long-range interactions, such as attraction \cite{park_BARWlongrange, park_ARWlongrange} and reaction \cite{odea_khemani_longrangeZ2absorbing}, were previously studied in $\mathbb{Z}_2$-symmetric models. This is in the spirit of active error-correction approaches in quantum computing, in which nonlocal information, namely, the global profile of error syndromes, is used to correct errors locally. Likewise, treating domain walls as the analogs of the syndromes, we obtain nonlocal classical information---the spatial distribution of the domains---which we then use to bias the walks, but all the applied gates are still local (two-site). We now show that such a model exhibits an absorbing state transition in a \textit{new} universality class that has not been reported to date.

We begin as before by first measuring the domain walls. Then, for each domain wall, we adopt a nonlocal rule with probability $q$, while with probability $1-q$, we follow the local rule introduced in the previous sections.  The specific nonlocal rule is as follows. Consider a single domain wall separating two domains, which we call $\mathcal{L}$ (on its left) and $\mathcal{R}$ (on its right). Then, we search the next-nearest domain to the left (right) of the wall and record the distance $r_\mathrm{left}$ ($r_\mathrm{right}$) if this domain is of the same type as $\mathcal{R}$ ($\mathcal{L}$). If the domain types are different, the corresponding distance is set to infinity. We then bias the wall's next step to drift one unit toward the closer domain, i.e., move left if $r_\mathrm{left}<r_\mathrm{right}$ and right if $r_\mathrm{left}>r_\mathrm{right}$. If the two domains are equidistant $(r_\mathrm{left}=r_\mathrm{right})$, the local rule is applied. In this way, our update makes use of 
nonlocal information while maintaining the locality of the applied gates. Note that this is, in a certain sense, minimally nonlocal: only information about the closest other domain walls to the left and the right is used.

The basic principle of our nonlocal rule is to enhance the chance of domains merging to become larger domains, thus leading to the absorbing state for sufficiently large $q$. 
Therefore, after biasing the walk with nonlocal information, the absorbing transition---that was absent in the three-state local model---becomes apparent.  This is illustrated in Fig.~\ref{fig:nonlocal}(c-e), which shows snapshots of each phase and the critical point. We find the critical exponents for the transition with nonlocal information to be
\fix{$\theta=0.30\pm0.02$, and $\nu_\parallel=0.26\pm0.03$}. These exponents are different from those of known directed percolation models, indicating a new universality class for this phase transition.

Nonlocal information can similarly be included in the dynamics of $\mathbb{Z}_2$-symmetric models, leading to extended regions of both absorbing and active phases in the phase diagram of parameters $p$ and $q$. The critical exponent along the critical line for $q>0$ is \fix{$\theta=0.34\pm0.01$}, which differs from that of the DP2 universality class observed at $q=0$. Moreover, the absorbing phase with nonlocal information is qualitatively distinct from that governed by purely local rules. In particular, the bubble lifetime and width distributions no longer exhibit tails that are inverse powers of time and width; instead, they follow exponentially decaying distributions characterized by finite correlation time and length scales (see details in the SM \cite{supp_mat}).

\section{Summary and Discussion.}
Starting with the simple case of a $\mathbb{Z}_2$-symmetric two-state model, in this work, we first revisited the paradigmatic absorbing state transition using a two-site feedback mechanism. By initializing the system with a single domain wall, we uncovered the universal dynamics of both the absorbing and active phases as well as the transition between them. Building on these results, we extended our analysis to three-state models and investigated the role of branching processes. The branching process is always relevant for local three-state models, with even infinitesimal branching rates inevitably driving the system into the active phase. Despite this, we showed how absorbing phases can still be achieved by incorporating nonlocal operations, which effectively counteract the branching-driven active phase.

Additionally, we propose using nonlocal information in combination with local feedback, drawing parallels to the LOCC framework in error correction. Unlike active decoding methods such as minimal-weight perfect matching~\cite{Dennis_topological_quantum_memory,higgott_MWPM}, which require full knowledge of syndrome distributions (akin to domain walls in our models), our nonlocal-information rule suggests that knowledge of a finite number of domain walls is sufficient to drive the system into the absorbing phase. This raises interesting open questions about the extent of nonlocal information necessary for decoding and the range of global information required.

An interesting application of this work would be to interpret the classical absorbing phase as a resource for passively protecting quantum information in architectures where one type of error (either bit-flip or phase-flip) dominates, e.g., cat codes~\cite{Leghtas2015_cat,Ofek2016_cat,Touzard2018_cat,Lescanne2020_cat,Berdou2023_cat,Reglade2024_cat,Putterman2025_cat}. This approach succeeds as long as the dissipation channel that implements the classical rules preserves the strong symmetry associated with the encoding of quantum information~\cite{albert2014,albert2016,Lieu_2020,Lieu_2024}.

\begin{acknowledgments}
\emph{Acknowledgments.}---We thank Sajant Anand,  Sarang Gopalakrishnan, \fix{Peter Grassberger}, Kyungjoo Noh, Nicholas O'Dea, Grace Sommers, and Yifan Zhang for useful discussions. This work was supported in part by NSF QLCI grant OMA-2120757. R.S. is supported by the Princeton Quantum Initiative Fellowship. 
\end{acknowledgments}

\onecolumn\newpage
\appendix

\section{Survival probability of $p$ random-walking particles}
\label{supp:survivalP}
In this section, we derive the survival probability of $p$ random walkers following the method introduced in~Refs.\cite{huse_fisher_CIC_prb1984, Fisher_walkers1984}. This survival probability is directly related to the relevance of  branching in the absorbing phase. Specifically, consider $p$ identical particles on a one-dimensional line, initially located at $x_1(0)<\cdots <x_p(0)$.  The survival probability at time $t$ is defined as the probability of random walkers proceeding without passing each other, maintaining the ordering 
\begin{align}
    \quad x_1(t')<\cdots <x_p(t'),\, \forall\, t'\in[0,t].
    \label{eqn:survival}
\end{align}
Each particle random walks with diffusivity $D$, and the associated microscopic model could, in principle, be described with an update rule on discrete space and time~\cite{huse_szpilka_fisher_latticemodel}. However, since we are interested in the long-time asymptotics, we consider the coarse-grained continuum limit of both the  time $t$ and  the location of the $i$-th particle, $x_i(t)$; this simply describes Brownian motion.
In $d$ dimensions (with infinite boundaries), the probability density of a free particle departing from $\vec{x}_i\in \mathbb{R}^d$ at time $t_0$ to be located at $\vec{x}_f$ at time $t_f=t_i+\Delta t$ in the continuum limit is 
\begin{align}
\label{eq:Green}
    G_d^0\left(\vec{x}^{}_f,t^{}_f \vert \vec{x}^{}_i,t^{}_i\right) 
    = \frac{1}{(2\pi D \Delta t)^{d/2}} e^{-\frac{\left|\vec{x}_f - \vec{x}_i\right|^2}{2D\Delta t}} 
    = \frac{1}{(2\pi D \Delta t)^{d/2}} e^{-\frac{\left(|\vec{x}_f|^2 + |\vec{x}_i|^2\right)}{2D\Delta t}}
    \exp\left(\vec{x}^{}_f \cdot \vec{x}^{}_i/D\Delta t\right),
\end{align}
which is the solution of the linear diffusion equation $\partial_t G - D\nabla^2 G = 0$ with an initial condition of $G(\vec{x},t_i) = \delta^d(\vec{x}-\vec{x}_i)$.

One can interpret a random walk of $p$ particles on a line as the random walk of a single particle in a $p$-dimensional space. The initial location of this single particle is $\vec{r} = (x_1(0),\cdots ,x_p(0)) \in \mathbb{R}^p$. Unlike the infinite boundary conditions assumed in the derivation of Eq.~\eqref{eq:Green} above, the noncrossing condition~\eqref{eqn:survival} imposes an \textit{absorbing} boundary condition, requiring the probability density to vanish at $x_1 = x_2,\cdots,x_{p-1}=x_p$.

The solution in this case can be found using the method of image charges since it is required to be zero on certain boundaries and the differential equation is linear. Specifically, we place the image charges at $\vec{r}_g\equiv (r_{g(1)},\cdots r_{g(p)}) = (x_{g(1)}(0),\cdots x_{g(p)}(0))$ where $g\in \mathbb{S}_p$ is a permutation of $p$ elements. The sign of each image charge is determined by $\mathrm{sgn}(g)$, which is $+1$ for even and $-1$ for odd permutations. Therefore, the solution of the diffusion equation is written as
\begin{align}
    G^{}_p(\vec{x},t) 
    &= \sum_{g\in \mathbb{S}_p} \mathrm{sgn}(g)~ G_p^0 (\vec{x},t|\vec{r}^{}_g,0) =\frac{1}{(2\pi D t)^{p/2}} e^{-\frac{|\vec{x}|^2 + |\vec{r}_g|^2}{2Dt}}
    \sum_{g\in \mathbb{S}_p} \mathrm{sgn}(g) \exp(\vec{x} \cdot \vec{r}^{}_g/Dt),
\label{eq:appendix_image}
\end{align}
where we used the fact that $|\vec{r}_g|^2=|\vec{r}|^2$. We emphasize that the total sum of the solution~\eqref{eq:appendix_image} over all space at a given time is not 1, unlike the usual probability distribution. Instead, the sum within the domain defined by the absorbing boundary is the survival probability $P_s(t)$:
\begin{align}
    P_s(t) = \int_{x_1<\cdots<x_p} d^p\vec{x}\, G_p(\vec{x},t)
    =\int_{-\infty}^{\infty} dx_1 \int_{x_1}^{\infty} dx_2 \cdots \int_{x_{p-1}}^{\infty} dx_p G_p(\vec{x},t).
    \label{eq:appendix_survival}
\end{align}
We first calculate the lowest-order nonzero contribution in Eq.~\eqref{eq:appendix_image}, and plug it into Eq.~\eqref{eq:appendix_survival} to find the asymptotic scaling of the survival probability.

To do so, we begin by evaluating $F_p(\vec{x},t;\vec{r})\equiv \sum_{g\in \mathbb{S}_p} \mathrm{sgn}(g)\exp(\vec{x}\cdot \vec{r}_g / Dt)$. 
We Taylor expand the exponential and find the lowest-order polynomial that is nonzero after summing over permutations:
\begin{align}
    F_p(\vec{x},t;\vec{r}) \nonumber 
    &\equiv \sum_{g\in \mathbb{S}_p} \mathrm{sgn}(g)\exp(\vec{x}\cdot \vec{r}_g / Dt)\nonumber  =\sum_{g\in \mathbb{S}_p} \mathrm{sgn}(g) \exp\left(\frac{1}{Dt} \sum_{i=1}^p x_i r_{g(i)}\right)\nonumber   =\sum_{g\in \mathbb{S}_p} \mathrm{sgn}(g) \prod_{i=1}^p \exp\left(x_i r_{g(i)}/Dt\right) \nonumber\\
    &= \sum_{\{n\}} 
    \left(
    \sum_{g\in \mathbb{S}_p} \mathrm{sgn}(g)
     \frac{1}{\prod_i n_i!  (Dt)^{\sum_i n_i}} \prod_{i=1}^p (x_ir_{g(i)})^{n_i}
    \right).
\end{align}
It is easy to show from Lemmas~\ref{lemma1} and \ref{lemma2} that the polynomials with a nonzero value after summing over permutations are \textit{totally nonsymmetric}. Therefore, they must have distinct exponents for each of the parameters $x_1,\cdots,x_p$, implying that the corresponding lowest order is $0+1+\cdots +(p-1) = \frac{p(p-1)}{2}$.
The lowest-order contribution after summing over permutations $g$ is precisely
\begin{align}
    (Dt)^{-\frac{p(p-1)}{2}}
    \frac{1}{0!1!\cdots(p-1)!}
    \sum_{h,g\in \mathbb{S}_p} \mathrm{sgn}(g) \prod_{i=1}^{p}\left(x_{h(i)} r_{g\cdot h(i)}\right)^{i-1},
\end{align}
which can be simplified using the relation:
\begin{align}
    \sum_{g\in \mathbb{S}_p} \mathrm{sgn}(g) \prod_{i=1}^{p}(r_{g\cdot h(i)})^{i-1} 
    &=\mathrm{sgn}(h) \sum_{g\in \mathbb{S}_p} \mathrm{sgn}(g) \prod_{i=1}^{p}(r_{g(i)})^{i-1} \nonumber\\
    &=\mathrm{sgn}(h)
    \begin{vmatrix}
    1 & r_1 & r_1^2 & \cdots & r_1^{p-1} \\ 
    1 & r_2 & r_2^2 & \cdots & r_2^{p-1} \\
    \vdots & \vdots & \vdots & \ddots &\vdots \\
    1 & r_p & r_p^2 & \cdots & r_p^{p-1} \\ 
    \end{vmatrix}
    =
    \mathrm{sgn}(h) \prod_{1\leq i <j\leq p} (r_j - r_i),
\end{align}
where we have substituted $g\rightarrow g\cdot h^{-1}$ and used $\mathrm{sgn}(h^{-1}) = \mathrm{sgn}(h)$ in the first equation. The final expression simplifies using the Vandermonde determinant.
Therefore, the lowest-order contribution to $F_p(\vec{x},t;\vec{r})$ is given by
\begin{align}
    F_p(\vec{x},t;\vec{r}) 
    &=
    (Dt)^{-\frac{p(p-1)}{2}}
    \frac{1}{1!\cdots(p-1)!}
    \prod_{1\leq i <j\leq p} (r_j - r_i)
    \sum_{h\in \mathbb{S}_p} 
    \mathrm{sgn}(h) \prod_{i=1}^p (x_{h(i)})^{i-1} + \cdots \nonumber\\
    &= (Dt)^{-\frac{p(p-1)}{2}}
    \frac{1}{1!\cdots(p-1)!}
    \prod_{1\leq i <j\leq p} (r_j - r_i)(x_j - x_i)
     + \cdots,
\label{eq:appendix_Fp}     
\end{align}
where we again use the Vandermonde determinant for $\vec{x}$. Finally, substituting  Eqs.~\eqref{eq:appendix_image} and \eqref{eq:appendix_Fp} into Eq.~\eqref{eq:appendix_survival}, and by rescaling $\vec{y}=\sqrt{2Dt}\vec{x}$, we find the survival probability
\begin{align}
    P_s(t)
    = 
    \left(
    2^{\frac{p(p-1)}{4}} \pi^{-\frac{p}{2}}
    \frac{1}{1!\cdots(p-1)!}
    \prod_{1\leq i <j\leq p} (r_j - r_i)
    ~\mathcal{I}_p
    \right) 
    (Dt)^{-\frac{p(p-1)}{4}}  e^{-\frac{|\vec{r}|^2}{2Dt}},
\end{align}
where
\begin{align}
    \mathcal{I}_p \equiv \int_{y_1<\cdots<y_p} d^p\vec{y}~ e^{-|\vec{y}|^2}  \prod_{1\leq i <j\leq p}(y_j - y_i)
\end{align}
is the Gaussian integral of the Vandermonde determinant, which is a function of $p$.

For large $t$, the survival probability, which is the probability that $p$ random-walking particles do not cross each other for at least time $t$, decays as
\begin{align}
    P_s(t) \sim t^{-\frac{p(p-1)}{4}}.
\end{align}
The survival probability is the cumulative of the lifetime probability distribution, $P_s(t) = \int^\infty_t P(\tau)d\tau$.
By differentiating the relation with respect to $t$, we obtain the asymptotic behavior of the lifetime probability distribution:
\begin{align}
    P(t) \sim t^{-\frac{p^2-p+4}{4}}.
\end{align}
Specifically, in the main text, we use $P(t) \sim t^{-3/2}$ for $p=2$ and $P(t) \sim t^{-5/2}$ for $p=3$ to demonstrate the relevance of the branching process at the absorbing fixed point.

Finally, we note here the lemmas used to derive the survival probabilities.
\begin{lemma}
For a p-variable function $f(x_1,\cdots,x_p)$, if there exists an odd permutation $\sigma$ such that
$f(x_1,x_2,\cdots x_p) = f(x_{\sigma(1)},x_{\sigma(2)},\cdots,x_{\sigma(p)})$, then
\begin{align}
    \sum_{g\in \mathbb{S}_p} \mathrm{sgn}(g) f(x_{g(1)},\cdots,x_{g(p)}) = 0.
\end{align}
\label{lemma1}
\end{lemma}
This is trivial because $\sigma \mathbb{S}_p = \mathbb{S}_p$, and $\mathrm{sgn}(\sigma)=-1$.

\begin{lemma}
    Consider a $p$-variable polynomial $x_1^{n_1}\cdots x_p^{n_p}$. If there exists $i\neq j$ such that $n_i = n_j$, then
    \begin{align}
    \sum_{g\in \mathbb{S}_p} \mathrm{sgn}(g) x_{g(1)}^{n_1}\cdots x_{g(p)}^{n_p} = 0
\end{align}
\label{lemma2}
\end{lemma}
This is also trivial using the lemma above, by considering an odd permutation $\sigma$ that permutes $i$ and $j$.

\section{Finite-size and time scaling for the absorbing state transition}
This section provides a brief review of finite-size and time scaling relations for absorbing state transitions and outlines the numerical methods used to determine the scaling exponents in our work. For a more comprehensive overview, we direct the reader to Refs.~\cite{Hinrichsen_absorbing_review, odor2004_nonequilibrium_review, henkel_absorbing_book}.

Classical nonequilibrium phase transitions are often characterized by four critical exponents: $\beta$, $\beta'$, $\nu_\parallel$, and $\nu_\perp$. The first two are the order parameter exponents (see below), while $\nu_\parallel$ and $\nu_\perp$ are the temporal and spatial correlation length exponents, respectively. Symmetry constraints can relate these exponents; for example, time-reversal symmetry in (1+1)D directed percolation leads to $\beta = \beta'$, and similar relations hold for the $\mathbb{Z}_2$-symmetric directed percolation universality class as well \cite{Mussawisade_1998_symmetryinBARW}. For a control parameter $p$ that tunes between active and absorbing phases, the stationary active phase has a density scaling as $\rho \sim |p - p_c|^\beta$, while the ultimate survival probability (probability of not reaching the absorbing state) scales as $P_\infty \sim |p - p_c|^{\beta'}$. In the field-theoretic sense, $\beta$ is often associated with creation processes, while $\beta'$ relates to annihilation processes.

\begin{figure}[t]
    \centering
    \includegraphics[width=\textwidth]{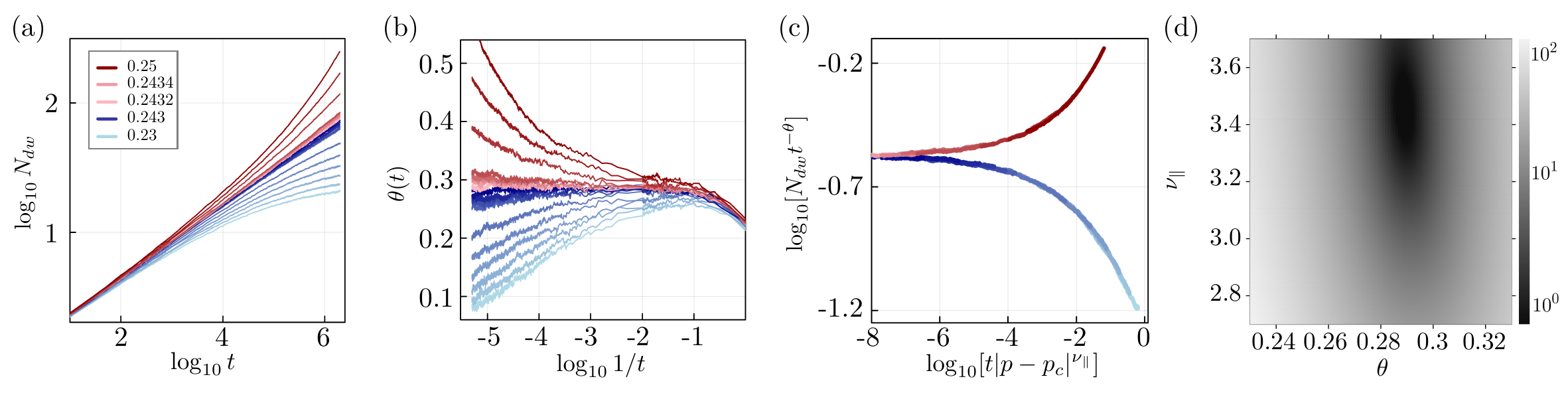}
    \caption{(a) Number of domain walls $N_{dw}$ in the local $\mathbb{Z}_2$-symmetric model over time plotted on a  log-log scale, for various branching rates $p$. (b) Effective exponent $\theta(t)$ as a function of $\log(1/t)$; the critical branching rate is when $\theta(t)$ saturates to a constant value as $\log (1/t)\rightarrow0$. (c) Data collapse after rescaling $N_{dw} t^{-\theta}$ against $t|q-q_c|^{\nu_\parallel}$ with $p_c\approx0.24$, $\theta\approx0.29$ and $\nu_\parallel\approx3.4$. (d) A density plot of $\log_{10} \mathcal{Q}$, where $\mathcal{Q}$ is a metric for the quality of the data collapse~\cite{scaling}, shows the optimal values of $\theta$ and $\nu_\parallel$ for the scaling ansatz. We average over $2\times 10^4$ samples.
    }
    \label{fig:Z2_critical}
\end{figure}

In our work, we initialize the system with a single particle---a domain wall---at the center of an infinite chain and track the particle count $N(t)$ over time. From this setup, we extract the time correlation length exponent $\nu_\parallel$ and a scaling exponent $\theta$, defined as 
\begin{align} 
\theta = \frac{d}{z} - \frac{\beta + \beta'}{\nu_\parallel}, 
\end{align}
where $\theta$ is derived from the connected function $\mathcal{C}(x,t)$, which describes the probability of finding a particle at position $x$ and time $t$, given an initial particle at $(x=0, t=0)$. The connected function follows the scaling form 
\begin{align} 
\mathcal{C}(x,t) \sim t^{\theta - d/z} f\left(\frac{x}{t^{1/z}}, |p - p_c|t^{1/\nu_\parallel}\right),
\end{align}
for some universal scaling function $f$, which leads to the relationship between $N(t)$ and $\theta$: 
\begin{align}
N(t) = \int d^d x ~ \mathcal{C}(x,t) \sim t^\theta \int d^d y ~ f\left(y,|p - p_c|t^{1/\nu_\parallel}\right)\sim t^\theta ~(\mathrm{when~}p=p_c). 
\end{align}
To estimate $\theta$, we calculate an effective exponent at a fixed parameter value: 
\begin{align}
\theta(t) = \log_{10} \frac{\bar{N}(10t)}{\bar{N}(t)}, 
\end{align}
where $\bar{N}$ is the particle number averaged over multiple samples.

Precisely at the critical point, the effective exponent $\theta(t)$ remains constant as $t \rightarrow \infty$. Away from $p_c$ (but still in its vicinity), the effective exponent will deviate from its critical value, curving upward or downward depending on whether the system is in the active or absorbing phase.

\section{Numerical results for the $Q=2$ local model}
In this section, we revisit the absorbing state transition in a model with $\mathbb{Z}_2$ symmetry, using a simpler setup to investigate the behavior deep within the active and absorbing phases. Our numerical results confirm the qualitative picture described in Figs.~1(b) and (c) of the main text.

\subsection{Critical point}

As described in the main text, this model involves only contiguous two-site gates at most, with the branching probability $p$ being the tuning parameter. Starting from an initial state with a single domain wall, we monitor the total number of domain walls $N_{dw}$ over time with infinite boundary conditions imposed.
Figure~\ref{fig:Z2_critical}(a) shows $N_{dw}(t)$ for various values of $p$, while Fig.~\ref{fig:Z2_critical}(b) displays the effective exponent $\theta(t)$, as defined in the previous section; our results indicate a critical exponent $\theta \approx 0.3$ and a critical branching rate $p_c \approx 0.24$. Using the scaling ansatz  $N_{dw}(t)\sim t^{\theta}f\left((p-p_c)t^{1/\nu_\parallel}\right)$ [Eq.~(1) from the main text], we find $\theta\approx0.29$ and $\nu_\parallel\approx3.4$, as shown in Fig.~\ref{fig:Z2_critical}(c) and (d). These values are consistent with the DP2 universality class.

\subsection{Active phase}
\begin{figure}
    \centering
    \includegraphics[width=\textwidth]{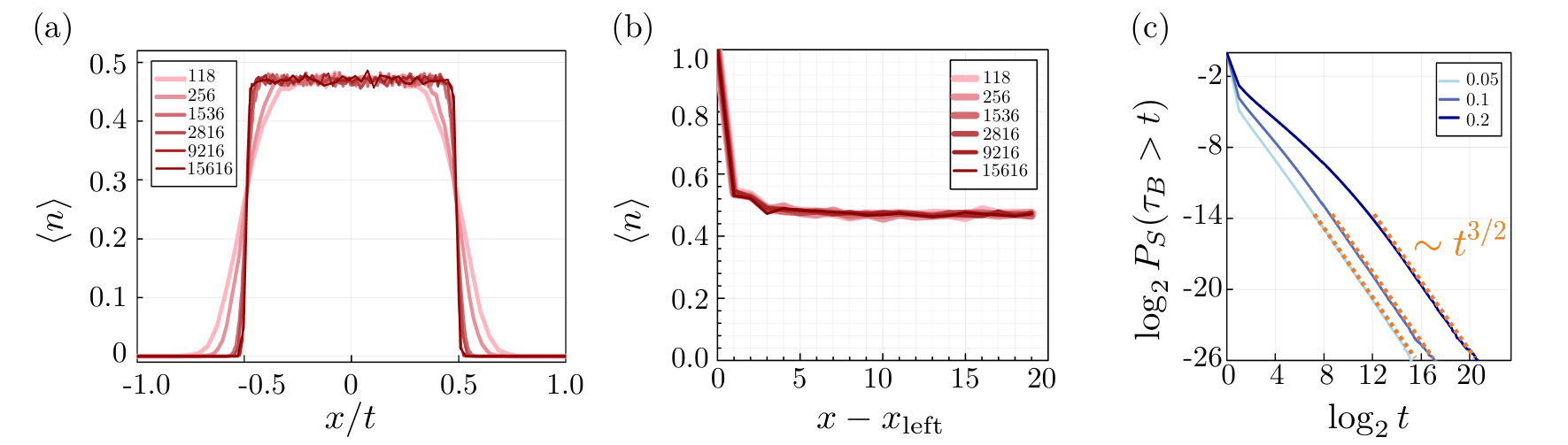}
    \caption{Numerical results for the local $\mathbb{Z}_2$-symmetric model in (a,b) the active phase with $p=0.6$, and (c) the absorbing phase with $p\in \{0.05,0.1,0.2\}$. (a) The number density profile $\langle n \rangle$ is displayed on a rescaled axis $x/t$ for various $t$ (labels), after averaging over $10^3$ samples.
    The perfect collapse demonstrates that the active domain initiated from a single domain wall ballistically spreads out, resulting in a uniform profile for the number density.
    (b) The profile near the edge of the ballistic front shows the sharp boundaries on the scale of $\mathcal{O}(1)$. The origin of the $x$-axis  is shifted by $x_{\mathrm{left}}$, the position of the leftmost site of the domain walls.
    (c) The cumulative distribution of the bubble lifetime showing an asymptotic tail scaling as $P_S(\tau_B>t)\sim t^{-3/2}$.
    }
    \label{fig:Z2_phases}
\end{figure}
In the active phase,  the system fails to reach an absorbing state within times that scale polynomially with the system size, since the branching is the dominant process. Therefore, even a single domain wall is sufficient to induce an active region that grows ballistically, developing a uniform domain-wall density within sharply defined boundaries.

To study the active phase, we choose a branching probability of $p=0.6$ (a value deep which puts us deep within the phase), and average over $10^4$ samples to obtain the spatial density profile over time. Figure~\ref{fig:Z2_phases}(a) depicts the spatial profile of the domain-wall density as a function of the rescaled variable $x/t$ at different times, averaged across samples. The density within the active region is uniform, while the edges expand ballistically, as evidenced by the collapse of the density profiles when plotted with the rescaled $x$-axis. Figure~\ref{fig:Z2_phases}(b) highlights the sharp boundaries of the active region,  which has a width $\sim \mathcal{O}(1)$.

\subsection{Absorbing phase}
The absorbing phase is governed by the annihilation fixed point in the renormalization-group sense, i.e., the diffusion and annihilation dominate and the branching becomes irrelevant \cite{Tauber_Cardy_prlPC,Tauber_Cardy_PC}. In 1+1 dimensions, starting from a random initial state of zeros and ones, the number of domain walls decays as $t^{-1/2}$. However, the scaling properties for an initial state with a single domain wall are not well understood. Here, we develop a framework to coarse-grain domain walls into \textit{bubbles} that behave like single particles.

When the branching probability $p$ is small, deep inside the absorbing phase, the annihilation dominates. A single domain-wall particle typically remains so without changing its number. Even if a particle branches occasionally, the domain wall is unlikely to proliferate as in the active phase and generally reverts to a single-particle state. We define the bubble lifetime $\tau_B$ as the duration in which a single particle branches, recombines, and returns to a single-particle state. The cumulative distribution of the bubble lifetime exhibits an asymptotic tail scaling as $P_s(\tau_B > t) \sim t^{-3/2}$ at large times [see Fig.~\ref{fig:Z2_phases}(c)]. This scaling exponent is derived by considering the survival time of three particles that perform random walks without colliding, which dominates the late-time behavior (see Sec.~\ref{supp:survivalP} for details). The asymptotic distribution of $\tau_B$ implies that the late-time dynamics are equivalent to waiting for three recombining bubbles. Bubbles essentially behave the same as single particles because the parity of the domain-wall number is the conserved quantity, and even a large bubble carries the same charge as the single particle from which it originates.

\section{Numerical results for $Q=3$ local models}
\label{subsection:S3_ex}
\begin{figure}
    \centering
    \includegraphics[width=\textwidth]{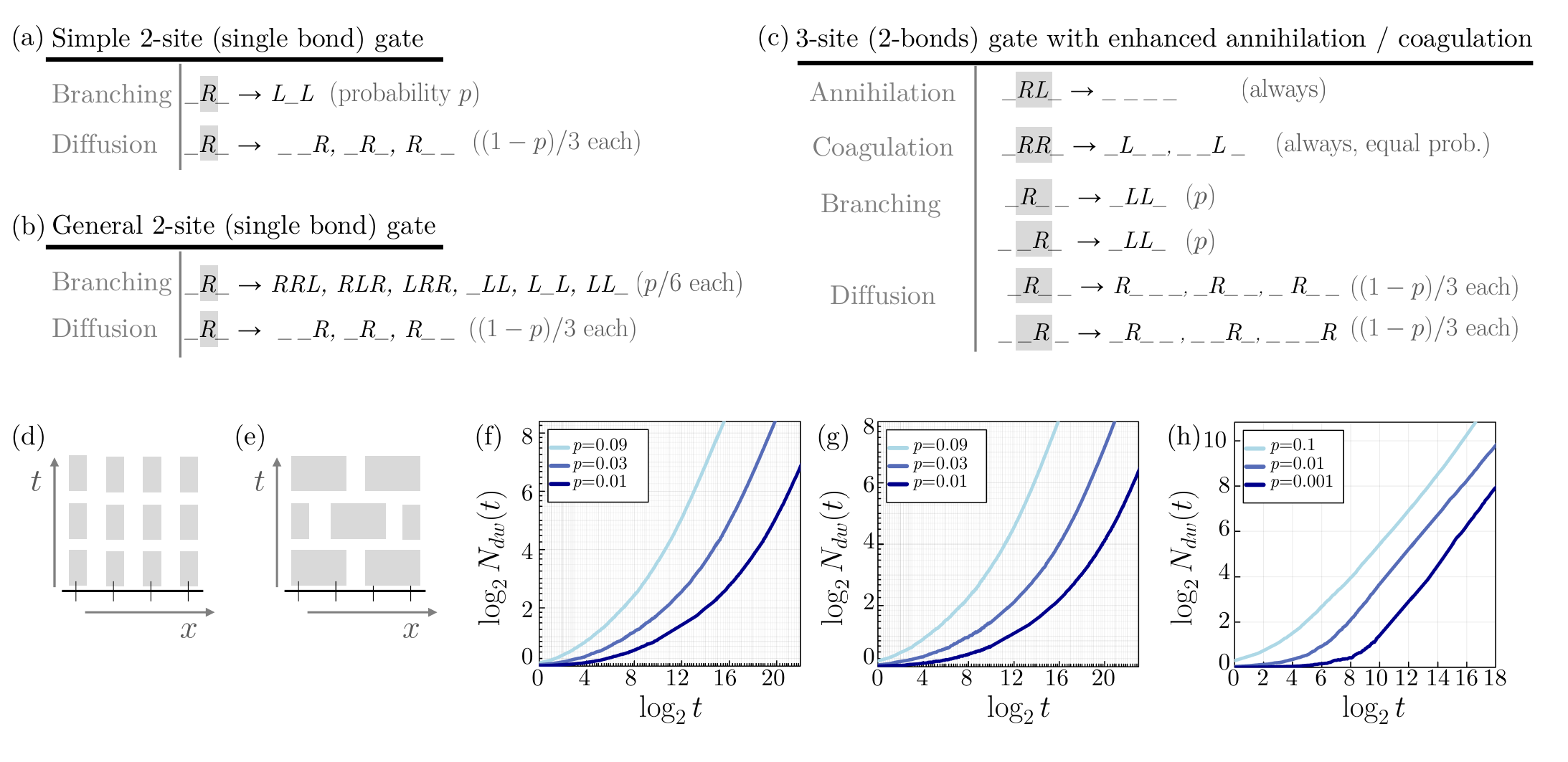}
    \caption{(a) Simple $Q=3$ model with two-site-gate feedback and a minimal branching process. Once a gate is applied to a single domain wall, it branches into two domain walls of different types with probability $p$. With probability $1-p$, the domain wall diffuses by hopping one unit left, one unit right, or staying in place with equal probabilities.
    (b) Generalized model with two-site-gate feedback. Unlike the previous model, a single domain wall can branch into six different possible configurations, all with equal probability. 
    (c) Feedback model with three-site gates acting on two adjacent bonds. The gate always annihilates two adjacent domain walls of different types and coagulates walls of the same type. If there is exactly one domain wall among two bonds, it branches into two domain walls of the other type with probability $p$ or performs a random walk with probability $1-p$.
    (d) Illustration of the synchronous operations for feedback rules (a) and (b), where gates are applied on every bond at each discrete time step.
    (e) The three-site gates are noncommutative, so they are applied in a brickwork pattern.
    (f), (g), and (h) show the total number of domain walls over time for rules (a), (b), and (c), respectively. All the $x$-axes are on a $\log_2$ scale. The number of domain walls always increases, even if the branching rate is small. }
    \label{fig:S3_numerics}
\end{figure}

We have shown that the branching process is always relevant to the annihilation fixed point for $Q=3$, and the system always becomes active, even for an infinitesimal branching rate. In this section, we numerically confirm this statement using specific models with two- and three-site gates.

The feedback models using two-site gates act on a single domain wall located on a given bond. As in the $\mathbb{Z}_2$-symmetric model introduced in the main text, we consider a space-time discrete setup with synchronous feedback [Fig.~\ref{fig:S3_numerics}(d)]. The domain walls branch with probability $p$, and diffuse with probability $1-p$. The detailed feedback rules are shown in Fig.~\ref{fig:S3_numerics}(a,b): the first rule corresponds to a minimal branching process in which $R\rightarrow 2L$, as described earlier in the main text, while the second rule allows branching into all the possible configurations, each with a probability of $p/6$, through the two-site gate operation.
Both these models lie in the active phase. Figures \ref{fig:S3_numerics}(f) and (g) show the total number of domain walls $N_{dw}(t) = n_R(t)+n_L(t)$, and the upward curvature of these plots indicates that the number of domain walls increases over time. On longer timescales, the curvature will flatten as $N_{dw}(t)\sim t$ in the asymptotic limit of the active phase.

We also confirm the relevance of the branching using longer-ranged gates, specifically, a three-site gate that acts on two bonds. To enhance the possibility of reducing the number of domain walls, we always apply pair annihilation or coagulation if the gate acts on two adjacent domain walls, depending on whether they are of the same type or not. Since such operations do not commute for two overlapping gates, we apply the feedback in a brickwork fashion, as illustrated in Fig.~\ref{fig:S3_numerics}(e). The detailed rules are shown in Fig.~\ref{fig:S3_numerics}(c). Likewise, the total domain-wall number increases with time [Fig.~\ref{fig:S3_numerics}(h)].
These numerical results support our earlier analytical findings, confirming that the system enters the active phase even for small branching probabilities.

\section{$\mathbb{Z}_2$-symmetric quantum circuits}
Consider a chain of $L$ qubits with open boundary conditions; our argument below holds for other boundary conditions too and the choice here is for convenience. We define the domain-wall operator
\begin{align}
    \hat{W}_{(r,r+1)} = \hat{Z}_r\hat{Z}_{r+1},\quad r\in \{1,\cdots, L-1\}.
\end{align}
Each domain-wall operator commutes with the others and has eigenvalues of $+1$ or $-1$, depending on whether a domain wall is present between two neighboring sites. The absorbing states for this system are $|\Uparrow\rangle = |\uparrow\rangle^{\otimes L}$ and $|\Downarrow\rangle = |\downarrow\rangle^{\otimes L}$, both of which are stabilized by the domain-wall operators:
\begin{align}
    \forall r,\quad \hat{W}_{(r,r+1)} |\psi\rangle = (+1)|\psi\rangle.
\end{align}
Although there are $L$ qubits, the set of independent domain-wall operators consists of only $L-1$ operators, leaving one remaining degree of freedom. This remaining degree of freedom corresponds to a logical bit representing the dimension of the Hilbert space we can effectively \textit{absorb}.  The following subsections present two types of quantum circuits equivalent to our classical reaction and diffusion rules. The first example is a stochastic interactive quantum circuit with projective measurements and probabilistic unitary feedback, while the second example considers a deterministic local channel.

\subsection{Description as a stochastic interactive quantum circuit}

We consider projective measurements of domain-wall operators $\hat{W}_{(r,r+1)}$ and apply feedback based on the measurement outcomes, as explained in the main text. If the measured value is $\hat{W}_{(r,r+1)} = +1$, no feedback is required, as there is no domain wall between the sites $r$ and $r+1$. However, if a domain wall is detected, indicated by $\hat{W}_{(r,r+1)} = -1$, the feedback is applied stochastically.

When we apply $\hat{X}_r$ as feedback, it anticommutes with both $\hat{W}_{(r,r+1)}$ and $\hat{W}_{(r-1,r)}$, effectively flipping their values (assuming $\hat{W}_{(r-1,r)}$ was also measured earlier and thus pinned to a certain value). This action shifts the domain wall from $(r,r+1)$ to $(r-1,r)$. Similarly, applying $\hat{X}_{r+1}$ shifts the domain wall to $(r+1,r+2)$. Therefore, the application of $\hat{X}_r$ and $\hat{X}_{r+1}$ with equal probability leads to an unbiased random walk or diffusion of the domain wall.
In addition to single-qubit feedback, we also consider applying $\hat{X}_r \hat{X}_{r+1}$. This operator commutes with $\hat{W}_{(r,r+1)}$ but anticommutes with both $\hat{W}_{(r-1,r)}$ and $\hat{W}_{(r+1,r+2)}$. As a result, it flips the two adjacent domain walls next to the original one, effectively creating a branching process $W\rightarrow 3W$.
    
For the synchronous update scheme introduced in the main text, all domain-wall operators are projectively measured simultaneously in the first layer of the circuit, fixing their values to $+1$ or $-1$. Each configuration is determined by the Born probability, effectively pinning the system's dynamics to those of classical domain walls encoded by $\langle \hat{W}_{(r,r+1)} \rangle$.

In the more general case, such as with asynchronous updates, domain-wall operators are measured stochastically, so they are not all pinned simultaneously. However, the typical timescale for each domain wall to be measured at least once scales as $\sim L \log L$, which is irrelevant compared to the classical absorbing timescale $\sim L^2$. Therefore, given any initial state, the interactive quantum circuit with projective measurements and subsequent feedback will always reduce to classical dynamics in the long term.

Therefore, on applying the classical absorbing  rule to any arbitrary quantum state (including mixed states), the system will eventually evolve towards the absorbing manifold, which is defined as a Bloch sphere and can be represented as
\begin{align} 
    \hat{\rho}^{}_{\mathrm{absorb}} = \hat{\mathbb{I}} + \vec{n}\cdot\vec{\tau}
\end{align}
where $|\vec{n}|\leq1$ and $\vec{\tau} = (\hat{\tau}_1,\hat{\tau}_2,\hat{\tau}_3)$ with
\begin{align}
    \hat{\tau}^{}_1 = |\Uparrow\rangle\langle\Downarrow| + |\Downarrow\rangle\langle\Uparrow|, 
    \quad
    \hat{\tau}^{}_2 = i(-|\Uparrow\rangle\langle\Downarrow| + |\Downarrow\rangle\langle\Uparrow|), 
    ~\text{and}\quad
    \hat{\tau}^{}_3 = |\Uparrow\rangle\langle\Uparrow| - |\Downarrow\rangle\langle\Downarrow|.
\end{align}
 The vector $\vec{n}$ is stochastically determined, depending primarily  on the initial state.

\subsection{Description as a deterministic quantum channel}
\label{sec:channel}

As argued in the previous section, certain stochastic $\mathbb{Z}_2$-symmetric interactive quantum circuits, with projective measurements and feedback, follow the classical reaction and diffusion rules that result in absorbing or active phases. If the parameters of the quantum circuit are in the absorbing phase, the evolved density matrix is described as a state within a Bloch sphere where the poles are $\vert\Uparrow\rangle\langle\Uparrow|$ and $|\Downarrow\rangle\langle\Downarrow|$. This is also true when we average the stochastic feedback into deterministic channels: since all the trajectories are absorbed into the Bloch sphere (which is a convex manifold), the averaged channel should also absorb to a state inside the Bloch sphere. Akin to the previous model---and as introduced in the main text---this channel has two steps which correspond to first measuring the domain walls and then applying the feedback. In order to have a well-defined channel, we also need an additional classical register that records the measurement data on each bond.

To describe the local channel acting on a bond between $r$ and $r+1$, we define the projectors,
\begin{align}
\label{eqn:2state_proj}
    \hat{\mathbf{P}}^{\emptyset}_{r,r+1} &= \frac{1}{2}\left(\hat{\mathbf{1}} + \hat{Z}_r\hat{Z}_{r+1} \right) = \frac{1}{2}\left(\hat{\mathbf{1}} +\hat{W}_{(r,r+1)}\right),\\
    \hat{\mathbf{P}}^{W}_{r,r+1} &= \frac{1}{2}\left(\hat{\mathbf{1}}- \hat{Z}_r\hat{Z}_{r+1} \right) = \frac{1}{2}\left(\hat{\mathbf{1}} -\hat{W}_{(r,r+1)}\right),
\end{align}
where $\hat{\mathbf{P}}^{\emptyset}_{r,r+1}$ and $\hat{\mathbf{P}}^{W}_{r,r+1}$ projects into the Hilbert space with a domain wall (between sites $r$ and $r+1$) absent or present, respectively. The first step of the channel is then described as
\begin{align}
    \mathcal{L}_{(r,r+1)}^{\text{step 1}}[\hat{\rho}] = \hat{\mathbf{P}}^{\emptyset}_{r,r+1} \hat{\rho} \hat{\mathbf{P}}^{\emptyset}_{r,r+1} \otimes |\emptyset\rangle\langle\emptyset| + \hat{\mathbf{P}}^{W}_{r,r+1} \hat{\rho} \hat{\mathbf{P}}^{W}_{r,r+1} \otimes |W\rangle\langle W|,
\end{align}
where $|\emptyset\rangle$ and $|W\rangle$ are orthonormal states representing the measurement output recorded in the classical register (corresponding to the absence or presence of a domain wall, respectively). Since all the domain-wall measurements commute with one another, the measurement channel on every bond could be applied simultaneously. After the measurement, we apply the second step of the channel:
\begin{align}
    \mathcal{L}^{\text{step 2}}_{(r,r+1)}[\hat{\rho}\otimes\eta] = \hat{\rho}\otimes \delta_{\eta,\emptyset} 
    ~+~
    \bigg(p\underbrace{\hat{X}_{r}\hat{X}_{r+1}\hat{\rho}\hat{X}_{r}\hat{X}_{r+1} }_{\text{branching}}
    ~+~ \frac{1-p}{3} \underbrace{\left(\hat{\rho}
    +  \hat{X}_{r}\hat{\rho}\hat{X}_{r}
    +  \hat{X}_{r+1}\hat{\rho}\hat{X}_{r+1}\right)}_{\text{diffusion (random walk)}} 
    \bigg)\otimes \delta_{\eta,W},
\end{align}
where $\eta$ denotes the register, and $\delta_{\eta,\emptyset}$ and $\delta_{\eta,W}$ represent the projection of the register onto $|\emptyset\rangle\langle\emptyset|$ or $|W\rangle\langle W|$. Again, the feedback channels commute with one another and could be applied simultaneously.

For a channel defined by averaging over the stochastic feedback and with a single absorbing state, \citet{odea_khemani_absorbingstate} argue that the dynamics of the diagonal term and the off-diagonal term of the density matrix are decoupled; hence to understand the dynamics of diagonal operators, e.g., $\hat{Z}$, it suffices to keep track of the diagonal elements which correspond to a classical Markov process. Similarly, in the models introduced in this paper---which have more than one absorbing state due to discrete symmetries---the projection operators $\hat{\mathbf{P}}^{\emptyset(W)}_{r,r+1}$ are decoupled from other degrees of freedom. In particular, the Heisenberg time evolution of the projectors is equivalent to the classical update rule on domain-wall particles, and one can determine whether the channel absorbs or not by investigating the classical reaction and diffusion process.

\section{$\mathbb{Z}_3$- and $\mathbb{S}_3$-symmetric quantum circuits}
We can also extend our discussion above to the three-state models. In this case, the following two-site projectors
\begin{align}
\label{eqn:3state_proj}
    \hat{\mathbf{P}}^0_{r,r+1} &= \left[|00\rangle \langle00| + |11\rangle \langle11| + |22\rangle \langle22|\right]_{r,r+1}, \\
    \hat{\mathbf{P}}^R_{r,r+1} &= \left[|01\rangle \langle01| + |12\rangle \langle12| + |20\rangle \langle20|\right]_{r,r+1}, \\
    \hat{\mathbf{P}}^L_{r,r+1} &= \left[|10\rangle \langle10| + |02\rangle \langle02| + |21\rangle \langle21|\right]_{r,r+1},
\end{align}
define either the absence of a domain wall or the existence of a domain wall of type $R$ or $L$. The three absorbing states are $|0\rangle^{\otimes L}$, $|1\rangle^{\otimes L}$, and $|2\rangle^{\otimes L}$, which satisfy
\begin{align}
    \forall r,\quad \hat{\mathbf{P}}_{r,r+1}^0 |\psi\rangle = |\psi\rangle.
\end{align}

After the projective measurement of the domain walls, we 
apply the feedback defined with single-site operators:
\begin{align}
\label{eqn:3state_flip}
    \hat{S}^R_r &= \left[|1\rangle\langle0| + |2\rangle \langle1| + |0\rangle \langle2|\right]_r, \\
    \hat{S}^L_r &= \left[|0\rangle\langle1| + |1\rangle \langle2| + |2\rangle \langle0|\right]_r,
\end{align}
where the two operators are related by conjugation $\hat{S}^R_r{} ^\dagger = \hat{S}_r^L$ and also commute with each other, $[\hat{S}_{r_1}^R,\hat{S}_{r_2}^L]=0$. If the state of the domain-wall variable is measured as $0$, $R$, or $L$, applying the feedback updates it to a different sector per the relations
\begin{alignat}{3}
    \hat{S}_r^R{}^\dagger \hat{\mathbf{P}}^0_{r,r+1} \hat{S}_r^R &= \hat{\mathbf{P}}^R_{r,r+1}
    ,\quad
    \hat{S}_r^R{}^\dagger \hat{\mathbf{P}}^R_{r,r+1} \hat{S}_r^R &&= \hat{\mathbf{P}}^L_{r,r+1}
    \text{, and}\quad
    \hat{S}_r^R{}^\dagger \hat{\mathbf{P}}^L_{r,r+1} \hat{S}_r^R &&= \hat{\mathbf{P}}^0_{r,r+1}, \\
    \hat{S}_r^L{}^\dagger \hat{\mathbf{P}}^0_{r,r+1} \hat{S}_r^L &= \hat{\mathbf{P}}^L_{r,r+1}
    ,\quad
    \hat{S}_r^L{}^\dagger \hat{\mathbf{P}}^R_{r,r+1} \hat{S}_r^L &&= \hat{\mathbf{P}}^0_{r,r+1}
    \text{, and}\quad
    \hat{S}_r^L{}^\dagger \hat{\mathbf{P}}^L_{r,r+1} \hat{S}_r^L &&= \hat{\mathbf{P}}^R_{r,r+1}.
\end{alignat}
Therefore, the dynamics reduce to classical rules being applied on the projected domain walls, and the same argument as before holds true. Similar to  Sec.~\ref{sec:channel}, this construction can also be trivially extended to quantum channels by generalizing the classical register recording the measurement outcome to $|0\rangle\langle0|$, $|1\rangle\langle1|$, and $|2\rangle\langle2|$.

\section{Transition with nonlocal information}
\subsection{Numerical results for $Q=3$ model}
We have demonstrated both theoretically and numerically that $Q > 2$ local models with discrete symmetry remain in the active phase, lacking a transition to the absorbing phase. However, introducing nonlocal information allows these models to also exhibit an absorbing phase transition. Specifically, we bias the dynamics by enforcing smaller domains to collapse while larger domains merge into even larger ones, using information about the length of the nearest domain (see the main text for detailed rules). At each discrete time step, the wall acquires nonlocal information with probability $q$, while with probability $1-q$, the system follows a locally defined $\mathbb{S}_3$ rule. The local rule that we choose is a random single-site flip adjacent to a measured domain wall. For example, a single domain wall $\textunderscore R\textunderscore$ can transition to one of $\textunderscore R\textunderscore$, $R\textunderscore~\textunderscore$, $\textunderscore ~\textunderscore R$, $\textunderscore LL$, or $LL \textunderscore$, each with equal probability $1/5$.

\begin{figure}
    \centering
    \includegraphics[width=\textwidth]{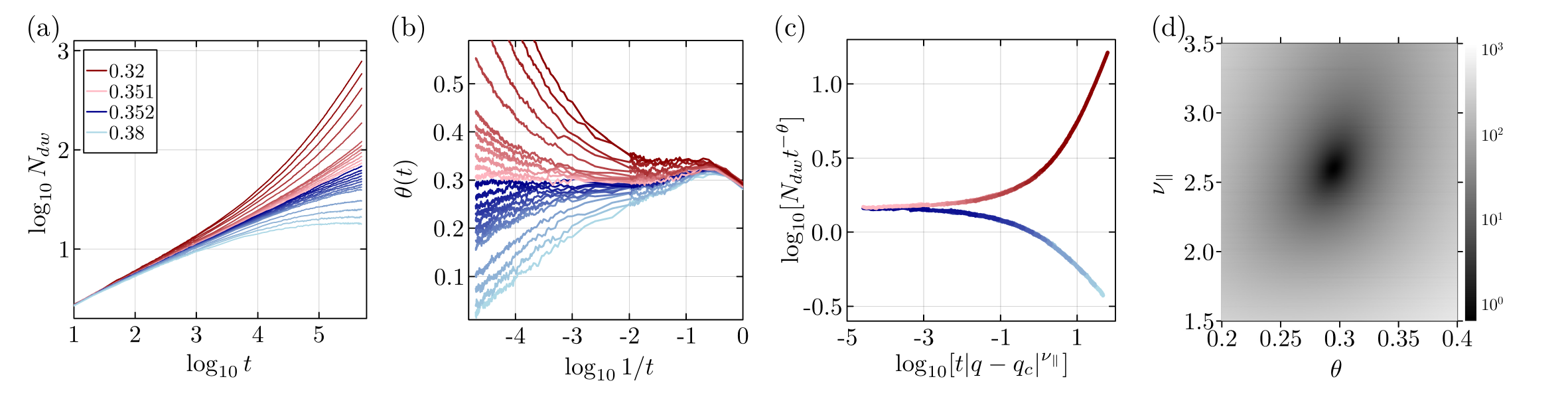}
    \caption{(a) Number of domain walls $N_{dw}$ in the $\mathbb{S}_3$-symmetric model over time plotted on a  log-log scale, for various rates of acquiring nonlocal information $q$. (b) Effective exponent $\theta(t)$ as a function of $\log(1/t)$; the critical branching rate is when $\theta(t)$ saturates to a constant value as $\log(1/t)\rightarrow0$. (c) Data collapse after rescaling $N_{dw} t^{-\theta}$ against $t|q-q_c|^{\nu_\parallel}$ with $q_c = 0.351$, $\theta = 0.295$, and $\nu_\parallel=2.560$. (d) A density plot of $\log_{10} \mathcal{Q}$, where $\mathcal{Q}$ is a metric for the quality of the data collapse~\cite{scaling}, shows the optimal values of $\theta$ and $\nu_\parallel$ for the scaling ansatz.
    }
    \label{fig:nonlocal_critical}
\end{figure}

To prevent undesired surface transitions, we start with two domain walls under periodic boundary conditions, ensuring that the walls are separated by a distance larger than any other characteristic length scale achievable within the runtime. We then monitor the evolution of the total number of domain walls, $N_{dw}$, originating from a single domain wall over time.
Figure~\ref{fig:nonlocal_critical}(a) shows $N_{dw}(t)$ for various values of $q$, while Fig.~\ref{fig:nonlocal_critical}(b) plots the effective exponent $\theta(t)$, suggesting $\theta \approx 0.3$ and $q_c \approx 0.35$. Using the scaling ansatz $N_{dw}(t)\sim t^{\theta}f\left((q-q_c)t^{1/\nu_\parallel}\right)$ as shown in Fig.~\ref{fig:nonlocal_critical}(c) and (d), we find $\theta\approx0.295$ and $\nu_\parallel\approx2.560$. Notably, these exponents differ from known directed percolation models, indicating a distinct universality class for the transition.

\subsection{Surface transition from a single-domain-wall initial state}
\begin{figure}
    \centering
    \includegraphics[width=0.9\textwidth]{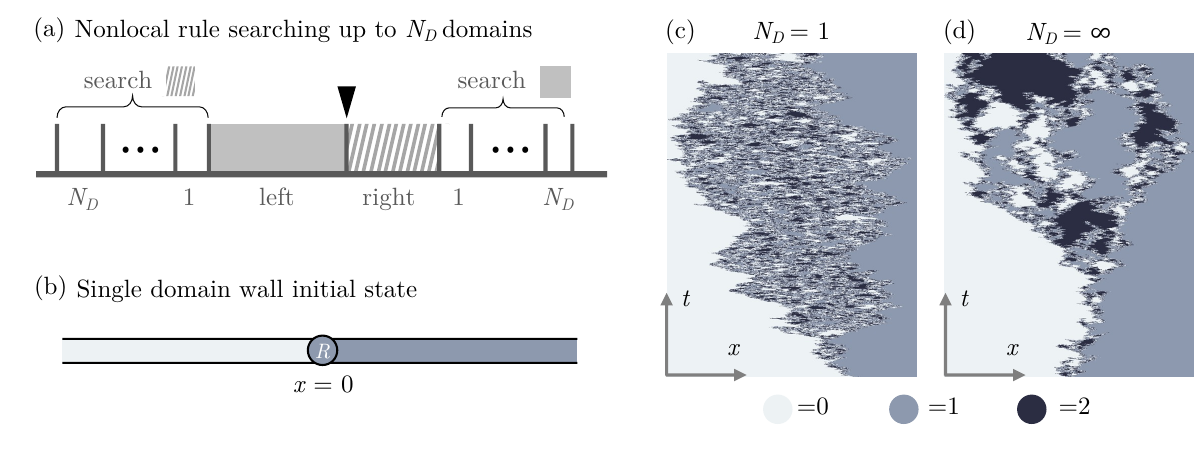}
    \caption{(a) Extended update rules using nonlocal information up to $N_D\geq1$ next-nearest-neighboring domains. The domain wall of interest separates a ``filled'' domain on the left from a ``striped'' domain on the right. Thus, we search for the striped domain within the next $N_D$ domains to the left and for a filled domain within the next $N_D$ domains on the right. (b) The initial state of a single domain wall in an infinite system yields a surface transition---different from the bulk transition---for such nonlocal actions.  Note that these are different boundary conditions from what is used in the main text. (c, d) Representative spatial configurations of domains at the critical point for $N_D = 1$ and $N_D=\infty$, as a function of time, underscoring the absence of an extended active region in the latter.
    }
    \label{fig:nonlocal_surface}
\end{figure}

As discussed in the main text, different boundary conditions and initial states can give rise to a surface transition distinct from the bulk transition. For instance, instead of initializing the system with two domain walls under periodic boundary conditions, starting with a single domain wall and infinite boundaries leads to qualitatively different dynamics. In this case, after a few branching events, the outermost domain walls cannot acquire information from one side due to the absence of additional domain walls. This asymmetry makes the outermost domain walls unique, resulting in a surface transition distinct from the bulk transition.

Considering the same nonlocal information described in the main text, the surface transition exhibits a first-order-like behavior. Specifically, at the critical point, four states coexist: three quiet (absorbing) domains of 0, 1, and 2,  along with an active domain characterized by small, short-lived clusters. The configurations at the critical point, as exemplified by Fig.~\ref{fig:nonlocal_surface}(c), clearly demonstrate the coexistence of active and quiet regions. This coexistence is suppressed if we extend the rule to search not only the next-nearest domain but up to $N_D\geq1$ nearest domains, as schematically shown in Fig.~\ref{fig:nonlocal_surface}(a). For instance, when $N_D=\infty$ as depicted in Fig.~\ref{fig:nonlocal_surface}(d), the active phase no longer coexists with the quiet domains at the critical point.

Additionally, we observe that the critical rate of acquiring nonlocal information, $q_c$, differs between the bulk and surface transitions. Naturally, the critical exponents for these transitions also differ, reflecting the distinct nature of the surface and bulk transitions.

For small $N_d$, in a scenario where neither of the domains (active or quiet) vanish, the boundaries between them will perform random walks. Consequently, the dynamical exponent should correspond to classical diffusion, with $z = 2$.
The critical exponent $\theta$ can be estimated by considering our earlier picture of the active phase for $Q=2$, where the number density of walls is uniform. The total number of walls proliferating from a single particle is proportional to the width of the active domain, leading then to $N(t) \sim t^\theta \sim t^{1/z}$, giving $\theta = 1/z = 1/2$.
Finally, to determine the temporal correlation length exponent $\nu_\parallel$, we can compare the two contributions to the motion of the boundary between the active and quiet domains. Over a given time $t$, the random walk driven by the local rules results in spatial fluctuation of the boundary proportional to $\sim\sqrt{t}$, while the nonlocal rules push the boundary towards the active domain by a distance $\sim(q-q_c) t$ since $q$ is the rate of nonlocal actions. The critical regime of this surface transition occurs when these competing effects are of similar magnitude, so the temporal correlation length scales as $t \sim (q - q_c)^{-2}$, yielding $\nu_\parallel = 2$. 

\subsection{$Q=2$ model with nonlocal information}
\begin{figure}
    \centering
    \includegraphics[width=\textwidth]{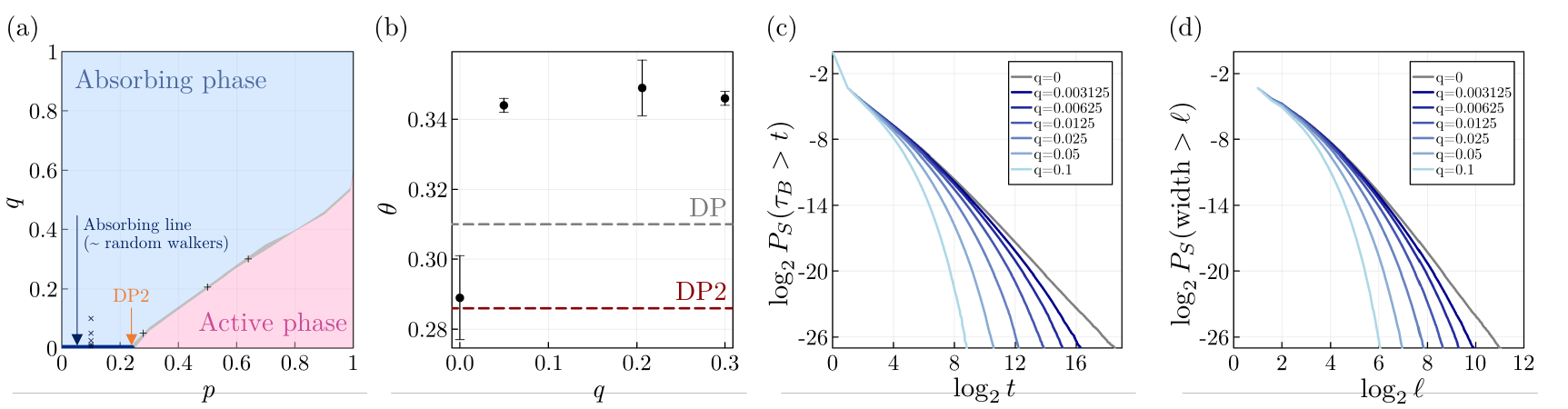}
    \caption{(a) Phase diagram in terms of branching rate $p$ and nonlocal information acquisition rate $q$ for the $\mathbb{Z}_2$-symmetric model.  As discussed in previous sections, when $q=0$, the transition between the absorbing and active phases occurs at $p_c\approx0.24$ and belongs to the DP2 universality class. For $q>0$, the absorbing and active phases extend, with $p_c(q)$ increasing as $q$ grows. However, the absorbing phase for $q>0$ is qualitatively distinct from the absorbing line at $q=0$, as described in the text. (b) The critical exponent $\theta$ for various $q$. At $q=0$, $\theta$ aligns with the DP2 universality class, whereas for $q>0$, it deviates from known directed percolation classes (DP and DP2, shown as dashed lines). These critical points are marked with crosses in (a). (c,d) Cumulative probability distributions of (c) bubble lifetimes and the (d) bubble widths, demonstrating that nonlocal information shifts the absorbing phase into a different fixed point from $q=0$. Both distributions are shown for $p=0.1$ (marked with x-crosses in (a)), which is well within the absorbing phase for $q=0$. When $q=0$, the distributions follow power-law decay: (c) $P_S(\tau_B>t)\sim t^{-3/2}$ and (d) $P_S(\mathrm{width}>\ell)\sim \ell^{-3}$. However, even an infinitesimally low rate $q>0$ bends the curve downward, leading to an exponential tail (not shown).}
    \label{fig:nonlocal_Z2}
\end{figure}
We have shown that nonlocal information can drive the active phase into the absorbing phase, as demonstrated in the $Q=3$ model. Here, we revisit this idea and apply it to $Q=2$ model. Unlike $Q=3$, where different domain wall types must be distinguished, the $Q=2$ model only has one domain wall type. As a result, the nonlocal information reduces to identifying the nearest domain walls on the left and right and determining which is closer. Additionaly, for $Q=2$, the behavior of a system initialized with two domain walls in a periodic boundary condition is equivalent to that of a system with a single domain wall in an infinite system.

As expected, introducing nonlocal interactions into the active phase of the $Q=2$ model stabilizes the system into an absorbing phase. The phase diagram is illustrated in Fig.~\ref{fig:nonlocal_Z2}(a). Interestingly, the critical exponents at the transition for $q>0$ differ from the DP2 universality class, which describes the transition at $q=0$, as shown in Fig.~\ref{fig:nonlocal_Z2}(b). Specifically, the critical exponent $\theta$ for $q>0$ is larger than the known values for the directed percolation (DP and DP2), marked with the dashed lines.

Beyond the critical point, the absorbing phase for finite $q$ is qualitatively distinct from the absorbing phase at $q=0$. As discussed in previous sections, the absorbing phase without nonlocal interactions can be coarse-grained into bubbles---which behave like random walking particles. This picture is supported by the cumulative probability distributions of the bubble lifetime (the duration before a single particle returns to a single particle after branching) and bubble width (the maximum separation between the left and right ends after a branching event and before recombination). As shown in the log-log plots of Fig.~\ref{fig:nonlocal_Z2}(c) and (d), deep in the absorbing phase without nonlocal information $(p,q)=(0.1,0)$, the distributions follow power-law scaling: 
\begin{align}
P_S(\tau_B>t)\sim t^{-3/2},\quad P_S(\mathrm{width}>\ell)\sim \ell^{-3}
\end{align}
where exponent 3 arises because the coarse-grained bubbles random walk with a dynamical exponent $z=2$. However, for finite $q$, the probability distributions develop an exponential tail:
\begin{align}
    P_S(\tau_B>t)\sim \exp(-t/\tau(p,q)), \quad P_S(\mathrm{width}>\ell)\sim \exp(-\ell/\ell(p,q))
    \label{eqn:exponential_tail}
\end{align}
introducing a correlation time $\tau(p,q)$ and correlation length $\ell(p,q)$. This demonstrates that the absorbing phase for $q>0$ is fundamentally different from the absorbing line at $q=0$, as illustrated in the phase diagram in Fig.~\ref{fig:nonlocal_Z2}(a).

\section{Connections to quantum error correction}
We now briefly remark on how the transition to an absorbing state, under certain circumstances, may be viewed as a form of error correction. Specifically, we consider the effect of a \textit{single layer} of bit-fip errors acting on the codeword and examine whether the system can be absorbed back into the initial codeword. We argue that the purely local circuits and those utilizing the nonlocal information exhibit distinct behaviors in this regard.

Suppose a small number of bit-flip errors occur, represented by the action of $\hat{X}_r$ on a few sites indexed by $r$. After these errors, the interactive quantum circuit with projective measurements and feedback will eventually put the system back into the absorbing manifold. If only $\mathcal{O}(1)$ bit-flip errors occur, the interactive circuit will restore the system to its initial wavefunction, resulting in $|\psi_f\rangle = |\psi_i\rangle$ with probability 1 in the limit $L \rightarrow \infty$.

However, consider a scenario where bit-flip errors occur at a finite \textit{density} $\gamma\ll1$, meaning a total of $\gamma L$ $\hat{X}$ operators are applied. In this case, while the interactive circuit still returns the system to the absorbing manifold, it may do so with a logical error, leading to the final state $|\psi_f\rangle = \hat{\tau}_1 |\psi_i\rangle = \beta|\Uparrow\rangle + \alpha|\Downarrow\rangle$. This failure occurs when random-walking domain walls are annihilated after crossing into a larger domain. 

To estimate the probability of this failure for the purely local circuit (corresponding to the absorbing line $q=0$ in the phase diagram Fig.~\ref{fig:nonlocal_Z2}(a)), we can model the problem as a single random-walking particle starting at position $x = \gamma L$, with absorbing boundaries at $x = 0$ and $x = L$. The probability that the particle reaches $x = L$ instead of $x = 0$ is $\gamma$. Consequently, the probability that the interactive circuit successfully restores the initial wavefunction, $|\psi_f\rangle = |\psi_i\rangle$, is $1 - \gamma$, while the probability of a logical $X$ error $|\psi_f\rangle = \hat{\tau}_1 |\psi_i\rangle$ is $\gamma$.

However, in the absorbing phase with a finite rate of nonlocal information ($q>0$), the circuit prioritizes eliminating the small domains, and the above argument is inapplicable. Instead, for small $p$ and nonzero $q$, the circuit introduces a characteristic length scale $\ell(p,q)$ [Eq.~\eqref{eqn:exponential_tail}]. Thus, for sufficiently large system size $L$, we expect the logical error rate will be exponentially suppressed as $\exp(-L/\ell(p,q))$. Note that this statement is for correcting one round of dilute spin flips.  This model is likely {\it not} fault-tolerant to spin flips that are occurring continuously in time at a nonzero local rate.

\bibliography{main}

\end{document}